# The Simultaneous Membership Problem for Chordal, Comparability and Permutation graphs

Krishnam Raju Jampani [*]     Anna Lubiw [†]

**Abstract**

In this paper we introduce the *simultaneous membership problem*, defined for any graph class $\mathcal{C}$ characterized in terms of representations, e.g. any class of intersection graphs. Two graphs $G_1$ and $G_2$, sharing some vertices $X$ (and the corresponding induced edges), are said to be *simultaneous members* of graph class $\mathcal{C}$, if there exist representations $R_1$ and $R_2$ of $G_1$ and $G_2$ that are "consistent" on $X$. Equivalently (for the classes $\mathcal{C}$ that we consider) there exist edges $E'$ between $G_1 - X$ and $G_2 - X$ such that $G_1 \cup G_2 \cup E'$ belongs to class $\mathcal{C}$.

Simultaneous membership problems have application in any situation where it is desirable to consistently represent two related graphs, for example: interval graphs capturing overlaps of DNA fragments of two similar organisms; or graphs connected in time, where one is an updated version of the other. Simultaneous membership problems are related to simultaneous planar embeddings, graph sandwich problems and probe graph recognition problems.

In this paper we give efficient algorithms for the simultaneous membership problem on chordal, comparability and permutation graphs. These results imply that graph sandwich problems for the above classes are tractable for an interesting special case: when the set of optional edges form a complete bipartite graph. Our results complement the recent polynomial time recognition algorithms for probe chordal, comparability, and permutation graphs, where the set of optional edges form a clique.

[*]David R. Cheriton School of Computer Science, University of Waterloo, Email:krjampan@uwaterloo.ca
[†]David R. Cheriton School of Computer Science, University of Waterloo, Email:alubiw@uwaterloo.ca

# 1 Introduction

We explore the idea of two graphs being *simultaneous members* of some graph class when the two graphs share some vertices and edges that must "behave consistently". We define this precisely for intersection graph classes, but the concept is rich enough to apply more broadly.

Let $\mathcal{C}$ be any intersection graph class (such as interval graphs, chordal graphs, permutation graphs, etc) and let $G_1$ and $G_2$ be two graphs in $\mathcal{C}$, sharing some vertices $X$ and the edges induced by $X$. $G_1$ and $G_2$ are said to be *simultaneous members of $\mathcal{C}$* or *simultaneous $\mathcal{C}$ graphs* if there exist intersection representations $R_1$ and $R_2$ of $G_1$ and $G_2$ such that any vertex of $X$ is represented by the same object in both $R_1$ and $R_2$. The *simultaneous membership problem* for class $\mathcal{C}$ asks whether $G_1$ and $G_2$ are simultaneous $\mathcal{C}$ graphs.

Comparability graphs do not have an intersection representation, but the simultaneous membership problem can be defined in the obvious way: Two comparability graphs $G_1$ and $G_2$ sharing some vertices $X$ and the edges induced by $X$ are said to be simultaneous comparability graphs if there exist transitive orientations $T_1$ and $T_2$ of $G_1$ and $G_2$ (respectively) such that any edge $e \in E(X)$ is oriented in the same way in both $T_1$ and $T_2$. For example, Figure 1(left) shows a pair of simultaneous comparability graphs, with the property that their union is not a comparability graph. Figure 1(right) shows a pair of graphs that are not simultaneous comparability graphs, though each one is a comparability graph on its own.

The main results in this paper are polynomial time algorithms for the simultaneous membership problem on chordal graphs, permutation graphs, and comparability graphs. These classes of graphs are of enduring interest because of their many applications [14, 8]. Simultaneous membership problems arise in any situation where two related graphs should be represented consistently. A main instance is for temporal relationships, where an old graph and a new graph share some common parts.

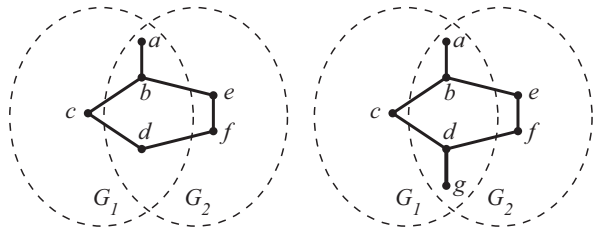

Figure 1: The graphs on the left are simultaneous comparability graphs while the graphs on the right are not.

Pairs of related graphs also arise in many other situations, for example: two social networks that share some members; overlap graphs of DNA fragments of two similar organisms, etc. Simultaneous chordal graphs have an application in computational biology as a special case of reconstructing phylogenies (tree structures that model genetic mutations) when part of the information is missing. (see [1]).

The simultaneous membership problem has previously been studied for straight-line planar graph drawings: two graphs that share some vertices and edges (not necessarily induced) have a *simultaneous geometric embedding* [3] if they have planar straight-line drawings in which the common vertices are represented by common points. Thus edges may cross, but only if they are in different graphs. Deciding if two graphs have a simultaneous geometric embedding is NP-hard [6].

Simultaneous membership problems are also closely related to some graph sandwich problems. For comparability graphs and for any intersection graph class we show that the simultaneous membership problem is equivalent to a graph augmentation problem: given two graphs $G_1$ and $G_2$, sharing vertices $X$ and the corresponding induced edges, do there exist edges $E'$ between $G_1 - X$ and $G_2 - X$ so that the augmented graph $G_1 \cup G_2 \cup E'$ belongs to class $\mathcal{C}$. Intuitively, the simultaneous membership problem does not specify relationships between $G_1 - X$ and $G_2 - X$, so these are the edges that can freely be added to produce a graph in class $\mathcal{C}$.

The *graph sandwich problem* [9] is a more general augmentation problem defined for any graph class $\mathcal{C}$: given graphs $G_1 = (V, E_1)$ and $G_2 = (V, E_2)$, is there a set of edges $E$ with $E_1 \subseteq E \subseteq E_2$ so that the graph $G = (V, E)$ belongs to class $\mathcal{C}$. This problem has a wealth of applications but is NP-complete for interval graphs, chordal graphs, comparability graphs, and permutation graphs [9].

The simultaneous membership problem is the special case where $E_2 - E_1$ forms a complete bipartite subgraph. A related special case where $E_2 - E_1$ forms a clique is the problem of recognizing *probe graphs*: a graph $G$ with a specified independent set $N$ is a *probe graph* for class $\mathcal{C}$ if there exist edges $E' \subseteq N \times N$ so that the augmented graph $G \cup E'$ belongs to class $\mathcal{C}$. Probe graphs have many applications [12, 10] and have received much attention recently. There are polynomial time algorithms to recognize probe interval graphs [11], probe chordal graphs [2], and probe comparability and permutation graphs [5].

Our paper is organized as follows: Section 1.1 gives notation and preliminaries. We prove the equivalence of the two problem formulations here. In sections 2, 3 and 4 we study the simultaneous membership problem for chordal, comparability and permutation graphs respectively.

## 1.1 Notation and Preliminaries

An *intersection graph* is one that has an intersection representation consisting of an object for each vertex such that there is an edge between two vertices if and only if the corresponding objects intersect. An intersection graph class restricts the possible objects, for example, interval graphs are intersection graphs of line segments on a line.

For a graph $G$, we use $V(G)$ and $E(G)$ to denote its vertex set and edge set respectively. Given a vertex $v$ and a set of edges $A$, we use $N_A(v)$ to denote the *neighbors of $v$* w.r.t A i.e. the vertex set $\{u : (u, v) \in A\}$. If $v$ is a vertex of $G$ then we use $E_G(v)$ to denote the edges incident to $v$ i.e. the edge set $\{(u, v) : u \in V(G), (u, v) \in E(G)\}$ and $G - v$ to denote the graph obtained by removing $v$ and its incident edges from $G$.

Let $G_1 = (V_1, E_1)$ and $G_2 = (V_2, E_2)$ be two graphs sharing some vertices $X$ and the edges induced by $X$. To be precise, $V_1 \cap V_2 = X$ and the subgraphs of $G_1$ and $G_2$ induced by vertex set $X$ are the same. Let $A \subseteq (V_1 - X) \times (V_2 - X)$ be a set of edges. We use the notation $(G_1, G_2, A)$ to denote the graph whose vertex set is $V_1 \cup V_2$ and whose edge set is $E_1 \cup E_2 \cup A$. Let $G = (G_1, G_2, A)$. An edge $e \in (V_1 - X) \times (V_2 - X)$ is said to be an *augmenting* edge of $G$.

**Theorem 1.** *Let $G_1 = (V_1, E_1)$ and $G_2 = (V_2, E_2)$ be two graphs belonging to intersection class $\mathcal{C}$ and sharing some vertices $X$ and the edges induced by $X$. $G_1$ and $G_2$ are simultaneous $\mathcal{C}$ graphs if and only if there exists a set $A \subseteq (V_1 - X) \times (V_2 - X)$ of edges such that the graph $G = (G_1, G_2, A)$ belongs to class $\mathcal{C}$.*

*Proof.* Let $G_1$ and $G_2$ be simultaneous members of $\mathcal{C}$. Let $R_1$ and $R_2$ be the intersection representations of $G_1$ and $G_2$ that are consistent on $X$. In this representation each vertex $v \in V_1 \cup V_2$ gets assigned to a set $T_v$. Now consider the intersection graph $G$ of $\{T_v : v \in V_1 \cup V_2\}$. Clearly $G$ belongs to class $\mathcal{C}$. Further $V(G) = V_1 \cup V_2$ and $E(G) = E_1 \cup E_2 \cup A$, where $A \subseteq (V_1 - X) \times (V_2 - X)$.

For the other direction suppose there exists a set $A \subseteq (V_1 - X) \times (V_2 - X)$ of edges such that the graph $G = (V_1 \cup V_2, E_1 \cup E_2 \cup A)$ belongs to class $\mathcal{C}$. Now consider the intersection representation $R$ of $G$. $R$ maps each vertex $v \in V_1 \cup V_2$ to a set $T_v$. For $i = 1, 2$, obtain a representation $R_i$ of $G_i$ by restricting the domain of $R$ to $V_i$. Note that $R_i$ is an intersection representation of $G_i$, since $G_i$ is the subgraph of $G$ induced on $V_i$. Now any vertex $v$ in $X$ is mapped to the same set ($T_v$) in both $R_1$ and $R_2$. Thus $G_1$ and $G_2$ are simultaneous members of $\mathcal{C}$. □

Theorem 1 implies that the simultaneous membership problem for intersection classes is a special case of the graph sandwich problem in which the set of optional edges induce a complete bipartite graph. In section 3, we show that this alternative formulation holds for comparability graphs also.

## 2 Simultaneous Chordal Graphs

A graph is said to be *chordal* if it does not contain any induced cycles of length greater than 3. We use the following well-known results about chordal graphs [8]. Chordal graphs satisfy the hereditary property: Any induced subgraph of a chordal graph is chordal. A chordal graph always has a *simplicial* vertex: a vertex $x$ such that $N(x)$ induces a clique. A *perfect elimination ordering* is an ordering $v_1, \ldots, v_n$ of the vertices such that each $v_i$ is simplicial in the subgraph induced by $\{v_i, \cdots, v_n\}$. Chordal graphs are characterized by the existence of a perfect elimination ordering. Any chordal graph is the intersection graph of a family of subtrees of a tree.

Let $G_1 = (V_1, E_1)$ and $G_2 = (V_2, E_2)$ be two chordal graphs sharing some common vertices $X = V_1 \cap V_2$ and the edges induced by $X$. Then (by Theorem 1), the simultaneous chordal graph problem asks whether there exists a set $A$ of augmenting edges such that the graph $(G_1, G_2, A)$ is chordal. We solve the following generalized problem: Given $G_1$, $G_2$ and $X$ (as above), and a set $F$ of *forced* augmenting edges, does there exist a set $A$ of augmenting edges such that the graph $(G_1, G_2, F \cup A)$ is chordal.

We need the following additional notation. For a vertex $v$ in $G = (G_1, G_2, F)$, we use $N_1(v)$ and $N_2(v)$ to denote the sets $N_{E(G)}(v) \cap V(G_1)$ and $N_{E(G)}(v) \cap V(G_2)$ respectively. Note that if $v \in V_1 - X$ then $N_2(v)$ may be non-empty because of $F$. Finally, we use $C(v)$ to denote the edge set $\{(x, y) : x \in N_1(v) - X, y \in N_2(v) - X\}$. A vertex $v \in G = (G_1, G_2, F)$ is said to be an *S-elimination vertex* of $G$ if $N_1(v)$ and $N_2(v)$ induce cliques in $G_1$ and $G_2$ respectively.

**Lemma 1.** *If $G = (G_1, G_2, F)$ is augmentable to a chordal graph then there exists an S-elimination vertex $v$ of $G$.*

*Proof.* Let $A$ be a set of augmenting edges such that the graph $G' = (G_1, G_2, F \cup A)$ is chordal. Because $G'$ is chordal it has a simplicial vertex, i.e. a vertex $v$ in such that $N_{E(G')}(v)$ induces a clique in $G'$. This in turn implies that $N_1(v)$ and $N_2(v)$ induce cliques. □

**Theorem 2.** *Let $G_1 = (V_1, E_1)$ and $G_2 = (V_2, E_2)$ be two graphs sharing some vertices $X$ and the edges induced by $X$. Let $G = (G_1, G_2, F)$ and let $v$ be any S-elimination vertex of $G$. $G$ is augmentable to a chordal graph if and only if the graph $G_v = (G_1, G_2, F \cup C(v)) - v$ is augmentable to a chordal graph.*

*Proof.* If $G_v$ is augmentable to a chordal graph, then there exists a set $A$ of augmenting edges such that $G'_v = (G_1, G_2, F \cup C(v) \cup A) - v$ is chordal. We claim that $G' = (G_1, G_2, F \cup C(v) \cup A)$ is chordal. Note that $N_{E(G')}(v) = N_1(v) \cup N_2(v)$, which forms a clique in $G'$. Thus $v$ is simplicial in $G'$. Furthermore, $G' - v = G'_v$ is chordal. This proves the claim. Thus $G$ can be augmented to a chordal graph by adding the edges $C(v) \cup A$.

To prove the other direction, assume without loss of generality that $v \in V_1$. Let $A$ be a set of augmenting edges of $G$ such that the graph $G' = (G_1, G_2, F \cup A)$ is chordal. Consider a subtree representation of $G'$. In this representation, each node $x \in V_1 \cup V_2$ is associated with a subtree $T_x$ and two nodes have an edge in $G'$ if and only if the corresponding subtrees intersect. We now alter the subtrees as follows.

For each node $x \in N_1(v) - X$ we replace $T_x$ with $T'_x = T_x \cup T_v$. Note that $T'_x$ is a (connected) tree since $T_x$ and $T_v$ intersect. Consider the chordal graph $G''$ defined by the (intersections of) resulting subtrees. Our goal is to show that the chordal graph $G'' - v$ is an augmentation of $G_v$, which will complete our proof. First note that $E(G'') \supseteq C(v)$ because for every $x \in N_1(v) - X$, subtree $T'_x$ intersects every subtree $T_y$ for $y \in N_{E(G')}(v)$. The only remaining thing is to show that the edges that are in $G''$ but not in $G'$ are augmenting edges, i.e. edges from $V_1 - X$ to $V_2 - X$. By construction, any edge added to $G''$ goes from some $x \in N_1(v) - X$ to some $y \in N_{E(G')}(v)$. Thus $x \in V_1 - X$, and we only need to show that $y \in V_2 - X$.

Note that $(y, v)$ is an edge of $E(G')$. Now if $y \in V_1$ then $(y, v) \in E_1$ and thus $x, y$ are both in the clique $N_1(v)$ and are already joined by an edge in $G$ (and hence $G'$). Therefore $y \in V_2 - X$ and we are done. □

Theorem 2 leads to the following algorithm for recognizing simultaneous chordal graphs

**Algorithm 1**
1. Let $G_1$ and $G_2$ be the input graphs and let $F = \{\}$.
2. **While** there exists an $S$-elimination vertex $v$ of $G = (G_1, G_2, F)$ **Do**
3. $\quad F \leftarrow F \cup C(v)$
4. $\quad$ Remove $v$ and its incident edges from $G_1, G_2, F$.
5. **End**
6. **If** $G$ is empty return YES **else** return NO

Note that the above algorithm also computes the augmented graph for the YES instances. We now show that Algorithm 1 can be implemented to run in time $O(n^3)$.

Determining whether a vertex $v$ is an $S$-elimination vertex is a key step of the algorithm. For this we have to check whether $N_1(v)$ and $N_2(v)$ induce cliques in $G_1$ and $G_2$ respectively. Note that although the sets $N_1(x)$ and $N_2(x)$ change as we add to the edge-set $F$, the graphs $G_1$ and $G_2$ are unchanged. The straightforward implementation takes $O(n^2)$ for this step. However we can improve this to $O(n)$ as explained below. In a chordal graph $H$ on $n$ nodes, given a set $X \subseteq V(H)$ of vertices, we can test whether $X$ induces a clique in $O(n)$ time as follows. Let $v_1, \cdots, v_n$ be a perfect elimination order of $H$ and let $v_i$ be the first vertex in this order that is present in $X$. Then $X$ induces a clique if and only if $N(v_i) \supseteq X$. This can clearly be tested in $O(n)$ time using adjacency matrices. Thus determining whether $v$ is an $S$-elimination vertex takes $O(n)$ time. Since we may have to check $O(n)$ vertices before finding an $S$-elimination vertex and since the number of iterations is $O(n)$, Algorithm 1 runs is $O(n^3)$ time.

## 3 Simultaneous Comparability Graphs

Recall that a *comparability graph* is defined as a graph whose edges can be transitively oriented. Golumbic [8] gave an $O(nm)$ time algorithm for recognizing comparability graphs and constructing the transitive orientation if it exists. In this section we extend Golumbic's [8] results to simultaneous comparability graphs and show that the simultaneous comparability problem can also be solved in $O(nm)$ time. We begin by proving the equivalence of the original definition of the simultaneous comparability graph problem and the augmenting edges version of the problem. This is the analogue of Theorem 1 which only applied to intersection classes.

We use the following additional notation. A directed edge from $u$ to $v$ is denoted by $\overrightarrow{uv}$. If $A$ is a directed set of edges, then we use $A^{-1}$ to denote the set of edges obtained by reversing the direction of each edge in $A$. We use $\hat{A}$ to denote the union of $A$ and $A^{-1}$. $A$ is said to be *transitive* if for any three vertices $a, b, c$, we have $\overrightarrow{ab} \in A$ and $\overrightarrow{bc} \in A \Rightarrow \overrightarrow{ac} \in A$. Our edge sets never include loops, so note the implication that if $A$ is transitive then it cannot contain a directed cycle and must satisfy $A \cap A^{-1} = \emptyset$ (because if it contained both $\overrightarrow{ab}$ and $\overrightarrow{ba}$ it would contain $\overrightarrow{aa}$). By definition a *transitive orientation* assigns a direction to each edge in such a way that the resulting set of directed edges is transitive. We use $G - A$ to denote the graph obtained by undirecting $A$ and removing it from graph $G$.

Let $G_1 = (V_1, E_1)$ and $G_2 = (V_2, E_2)$ be two comparability graphs sharing some vertices $X$ and the edges induced by $X$. If $G_1$ and $G_2$ are simultaneous comparability graphs, then there exist transitive orientations $T_1$ and $T_2$ of $G_1$ and $G_2$ (respectively) that are consistent on $E(X)$. We call $T = T_1 \cup T_2$ a *pseudo-transitive orientation* of $G_1 \cup G_2$. Note that the orientation induced by $V_1$ (and $V_2$) in $T$ is transitive. If $W \subseteq \hat{E}_1 \cup \hat{E}_2$, then $W$ is said to be *pseudo-transitive* if $W \cap \hat{E}_1$ and $W \cap \hat{E}_2$ are both transitive. We can show that any pseudo-transitive orientation of $G_1 \cup G_2$ can be augmented to a transitive orientation, which is the main ingredient in the proof of the following equivalence theorem.

**Theorem 3.** *Let $G_1 = (V_1, E_1)$ and $G_2 = (V_2, E_2)$ be two comparability graphs sharing some vertices $X$ and the edges induced by $X$. $G_1$ and $G_2$ are simultaneous comparability graphs if and only if there exists a set $A \subseteq (V_1 - X) \times (V_2 - X)$ of edges such that the graph $G = (V_1 \cup V_2, E_1 \cup E_2 \cup A)$ is a comparability graph.*

*Proof.* Let $A \subseteq (V_1 - X) \times (V_2 - X)$ be a set of edges such that the graph $G = (V_1, V_2, E_1 \cup E_2 \cup A)$ is a comparability graph. Let $T$ be a transitive orientation of $G$. For $i = 1, 2$ let $T_i$ be a (directed) subgraph of $T$ induced by $V_i$. Clearly $T_i$ is a transitive orientation of $G_i$. Further any edge in $E(X)$ gets the same orientation in both $T_1$ and $T_2$. Hence $G_1$ and $G_2$ are simultaneous comparability graphs.

For the other direction, let $G_1$ and $G_2$ be simultaneous comparability graphs. Let $T_1$ and $T_2$ be transitive orientations of $G_1$ and $G_2$ (respectively) that are consistent on $E(X)$. Then $T = T_1 \cup T_2$ is a pseudo-transitive orientation of $G_1 \cup G_2$. We now extend $T$ to a transitive orientation $T'$ by adding a set $A'$ of (directed) augmenting edges. We define $A'$ as follows

1. For all vertex triples $a, b, c$ with $a \in V_1 - X$, $b \in X$ and $C \in V_2 - X$, $\overrightarrow{ab} \in T$ and $\overrightarrow{bc} \in T \Rightarrow \overrightarrow{ac} \in A'$.

2. For all vertex triples $a, b, c$ with $a \in V_2 - X$, $b \in X$ and $c \in V_1 - X$, $\overrightarrow{ab} \in T$ and $\overrightarrow{bc} \in T \Rightarrow \overrightarrow{ac} \in A'$.

Now it is sufficient to prove that $T' = T \cup A'$ is transitive.

We first show that for any two vertices $a, c \in V_1 \cup V_2$ exactly one of $\overrightarrow{ac}$ and $\overrightarrow{ca}$ can be in $T'$. Suppose both $\overrightarrow{ac} \in T'$ and $\overrightarrow{ca} \in T'$ hold. Since $T$ is pseudo-transitive, $\overrightarrow{ac}$ and $\overrightarrow{ca}$ cannot both belong to $T$. Suppose $\overrightarrow{ac} \in A'$ with $a \in V_1 - S$ and $c \in V_2 - S$. Then $\overrightarrow{ca}$ must be in $A'$ as well (not in $T$). Thus (by definition of $A'$) there exist vertices $b, d \in X$ such that $\overrightarrow{ab} \in T$, $\overrightarrow{bc} \in T$, $\overrightarrow{cd} \in T$ and $\overrightarrow{da} \in T$. Now $b, a, d \in V_1$, therefore $\overrightarrow{da} \in T$ and $\overrightarrow{ab} \in T$ implies that $\overrightarrow{db} \in T$. Similarly $b, c, d \in V_2$, therefore $\overrightarrow{bc} \in T$ and $\overrightarrow{cd} \in T$ implies that $\overrightarrow{bd} \in T$. Thus $T$ contains both $\overrightarrow{bd}$ and $\overrightarrow{db}$ which contradicts that $T$ is pseudo-transitive.

Now let $\overrightarrow{ab}$ and $\overrightarrow{bc}$ belong to $T'$. It is enough to show that $\overrightarrow{ac} \in T'$.
Case 1: $\overrightarrow{ab} \in T$ and $\overrightarrow{bc} \in T$
Assume wlog that $a, b \in V_1$. If $c \in V_1$ then by transitivity of $T_1$, $\overrightarrow{ac} \in T_1 \subseteq T'$. Otherwise $c \in V_2 - X$, which forces $b \in X$, so by definition of $A'$, $\overrightarrow{ac} \in A' \subseteq T'$.
Case 2: $\overrightarrow{ab} \in T$ and $\overrightarrow{bc} \in A'$
Since $\overrightarrow{bc} \in A'$, we can assume without loss of generality that $b \in V_1 - X$ and $c \in V_2 - X$. Also by definition of $A'$, there exists a vertex $d \in X$ such that $\overrightarrow{bd} \in T$ and $\overrightarrow{dc} \in T$. Now $\overrightarrow{ab} \in T$ implies that $a \in V_1$ (since $b \in V_1$) and thus $a, b, d$ all belong to $V_1$. Since $\overrightarrow{ab} \in T$ and $\overrightarrow{bd} \in T$ we must have $\overrightarrow{ad} \in T$. Now if $a \in V_1 - X$ then $\overrightarrow{ac} \in A' \subseteq T'$ and if $a \in X$ then $ac \in T \subseteq T'$ (since $\{a, d, c\} \subseteq V_2$)
Case 3: $\overrightarrow{ab} \in A'$ and $\overrightarrow{bc} \in T$
This case is identical to case 3.
Case 4: $\overrightarrow{ab} \in A'$ and $\overrightarrow{bc} \in A'$
We can assume without loss of generality that $a, c \in V_1 - X$ and $b \in V_2 - X$.

Now $\overrightarrow{ab} \in A'$ implies that there exists a vertex $d \in X$ such that $\overrightarrow{ad} \in T$ and $\overrightarrow{db} \in T$
Similarly $\overrightarrow{bc} \in A'$ implies that there exists a vertex $e \in X$ such that $\overrightarrow{be} \in T$ and $\overrightarrow{ec} \in T$

Since $\vec{db}, \vec{be} \in T$ and $\{d, b, e\} \in V_2$, $\vec{de} \in T$. Now $a, d, e, c$ all belong to $V_1$, hence $\{\vec{ad}, \vec{de}, \vec{ec}\} \subseteq T$ implies $\vec{ac} \in T \subseteq T'$.

Thus in all cases $\vec{ac} \in T'$. Hence we conclude that $T'$ is transitive. □

We now sketch a high level overview of Golumbic's algorithm for recognizing comparability graphs and compare it with our approach. Golumbic's recognition algorithm is conceptually quite simple: orient one edge (call it a "seed" edge), and follow implications to orient further edges. If this process results in an edge being oriented both forwards and backwards, the input graph is rejected. Otherwise, when there are no further implications, the set of oriented edges (called an "implication class") is removed and the process repeats with the remaining graph. The correctness proof is not so simple, requiring an analysis of implication classes, and of how deleting one implication class changes other implication classes. Golumbic proves the following theorem.

**Theorem 4.** *(Golumbic [8]) Let $G = (V, E)$ be an undirected graph and let $\hat{E}(G) = \hat{B}_1 + \hat{B}_2 + \cdots \hat{B}_j$ be any "G-decomposition" where $B_k$ is an implication class of $G - \cup_{1 \leq l < k} \hat{B}_l$. The following statements are equivalent:*

1. *$G$ is a comparability graph.*

2. *$I \cap I^{-1} = \emptyset$ for all implication classes $I$ of $G$.*

3. *$B_k \cap B_k^{-1} = \emptyset$ for $k = 1, \cdots, j$.*

We follow a similar strategy except that the "seed" edges must be chosen carefully for our proof to work. We define the concept of a "composite class" which is analogous to an implication class. We further classify a composite class as a "base class" or a "super class" depending on whether it is disjoint from $E(G_1) \cap E(G_2)$ or not. Our algorithm works as follows: As long as there is a base class remove it and recursively orient the remaining graph. Otherwise (when there are no base classes left) as long as there is a super class remove it and recursively orient the remaining graph.

We prove the following theorem.

**Theorem 5.** *Let $G_1 = (V_1, E_1)$ and $G_2 = (V_2, E_2)$ be two comparability graphs sharing some vertices $X$ and the edges induced by $X$. Let $\hat{E}_1 \cup \hat{E}_2 = \hat{B}_1 + \hat{B}_2 + \cdots + \hat{B}_i + \hat{S}_{i+1} + \hat{S}_{i+2} + \cdots + \hat{S}_j$ be a "S-decomposition" of $G_1 \cup G_2$ where $B_k$ is a base class of $G - \cup_{1 \leq l < k} \hat{B}_l$ and $S_k$ is a super class of $G - \cup_{1 \leq l \leq i} \hat{B}_l - \cup_{i+1 \leq l < k} \hat{S}_l$*
*The following statements are equivalent.*

1. *$G_1$ and $G_2$ are simultaneous comparability graphs*

2. *$C \cap C^{-1} = \emptyset$ for all composite classes $C$ of $G = G_1 \cup G_2$.*

3. *$B_k \cap B_k^{-1} = \emptyset$ for $k = 1, \cdots, i$ and $S_k \cap S_k^{-1} = \emptyset$ for $k = i+1, \cdots, j$.*

We now formalize and justify the above defined notions. Given an undirected graph $H$, we can replace each undirected edge $(u, v)$ by two directed edges $\vec{uv}$ and $\vec{vu}$ and define a relation $\Gamma$ on the directed edges as follows. $\vec{ij} \Gamma \vec{i'j'}$ if ($i = i'$ and $(j, j') \notin E(H)$) or ($j = j'$ and $(i, i') \notin E(H)$). $\Gamma$ can be viewed as a constraint that directs the $(i, j)$ edge from $i$ to $j$ if and only if the edge $(i', j')$ is directed from $i'$ to $j'$. It is easy to see that the transitive closure of $\Gamma$, denoted by $\Gamma_t$, is an equivalence relation. We refer to the partitions of $\Gamma_t$ as *implication classes*. The following Lemmas capture some of the fundamental properties of implication classes.

**Lemma 2.** ([8]) Let $A$ be an implication class of a graph $H$. If $H$ has a transitive orientation $F$, then either $F \cap \hat{A} = A$ or $F \cap \hat{A} = A^{-1}$ and in either case, $A \cap A^{-1} = \emptyset$.

**Lemma 3.** ([8]) Let the vertices $a, b, c$ induce a triangle in $H$ and let $\overrightarrow{bc}, \overrightarrow{ca}$ and $\overrightarrow{ba}$ belong to implication classes $A, B$ and $C$ respectively. If $A \neq C$ and $A \neq B^{-1}$, then

1. If $\overrightarrow{b'c'} \in A$ then $\overrightarrow{b'a} \in C$ and $\overrightarrow{c'a} \in B$

2. No edge of $A$ is incident with $a$.

**Lemma 4.** ([8]) Let $A$ be an implication class of a graph $H$. If $A \cap A^{-1} = \emptyset$, then $A$ is transitive.

Note that in Lemma 2, if the directions of one or more edges of triangle $abc$ are reversed, then the Lemma can still be applied by inversing the corresponding implication classes. For e.g when $\overrightarrow{ab} \in C$, $\overrightarrow{ac} \in B$ and $\overrightarrow{bc} \in A$, if $A \neq C^{-1}$ and $A \neq B$, then condition (1) becomes: If $\overrightarrow{b'c'} \in A$ then $\overrightarrow{ab'} \in C$ and $\overrightarrow{ac'} \in B$.

Let $G_1 = (V_1, E_1)$ and $G_2 = (V_2, E_2)$ be two comparability graphs sharing some vertices $X$ and the edges induced by $X$. Let $G = G_1 \cup G_2$. We define a relation $\Gamma'$ on the (directed) edges of $G$ as follows: $\overrightarrow{e}\Gamma'\overrightarrow{f}$ if $\overrightarrow{e}\Gamma\overrightarrow{f}$ and $\{\overrightarrow{e}, \overrightarrow{f}\} \subseteq \hat{E}_1$ or $\{\overrightarrow{e}, \overrightarrow{f}\} \subseteq \hat{E}_2$. It is easy to see that the transitive closure of $\Gamma'$ denoted by $\Gamma'_t$ is an equivalence relation. We refer to the partitions of $\Gamma'_t$ as "composite classes". A composite class $C$ is said to be *pseudo-transitive* if $(\overrightarrow{ab} \in C$ and $\overrightarrow{bc} \in C) \Rightarrow \overrightarrow{ac} \in C$ whenever $\{a, b, c\} \in V_1$ or $\{a, b, c\} \in V_2$.

From the definition it follows that each composite class is a union of one or more of the implication classes of $G_1$ and the implication classes of $G_2$. If a composite class $C$ of $G$ has an edge that belongs to $E(X)$, then we use the term "super class" to refer to $C$. Otherwise $C$ is said to be a "base class". Thus any base class is a single implication class of $G_1$ or $G_2$ and is contained in $\hat{E}_1 - \hat{E}(X)$ or $\hat{E}_2 - \hat{E}(X)$.

**Observation:** Note that every implication class of a super class contains an edge $\overrightarrow{e} \in \hat{E}(X)$.

The following Lemmas for composite classes are analogous to Lemmas 2, 3 and 4.

**Lemma 5.** Let $A$ be a composite class of $G = G_1 \cup G_2$. If $F$ is a pseudo-transitive orientation of $G$ then either $F \cap \hat{A} = A$ or $F \cap \hat{A} = A^{-1}$ and in either case, $A \cap A^{-1} = \emptyset$.

*Proof.* Let $Y$ be a set of directed augmenting edges of $G = G_1 \cup G_2$ such that $F' = F + Y$ is a transitive orientation. Let $G'$ be the graph obtained by undirecting $F'$. Thus $G'$ is an augmentation of $G$. Now any composite class of $G$ is contained in some implication class of $G'$. Let $A'$ be the implication class of $G'$ that contains $A$. Note that $A \subseteq A' - Y$. Now the lemma follows by applying Lemma 2 on $A'$ and $G'$. □

**Lemma 6.** Let the vertices $a \in X$, $b$ and $c$ induce a triangle in $G = G_1 \cup G_2$, such that $\overrightarrow{bc}, \overrightarrow{ca}$ and $\overrightarrow{ba}$ belong to composite classes $A, B$ and $C$ respectively. If $A \neq C$ and $A \neq B^{-1}$, then

1. If $\overrightarrow{b'c'} \in A$ then $\overrightarrow{b'a} \in C$ and $\overrightarrow{c'a} \in B$.

2. No edge of $A$ is incident with $a$.

*Proof.* We cannot appeal immediately to Lemma 3 because although $\overrightarrow{bc}$ and $\overrightarrow{b'c'}$ belong to the same composite class $A$, they need not be in the same implication class.

Since $\overrightarrow{bc}, \overrightarrow{b'c'} \in A$, there exist a sequence of edges $\overrightarrow{b_1c_1}, \cdots, \overrightarrow{b_kc_k}$, such that $\overrightarrow{bc}\Gamma'\overrightarrow{b_1c_1}\Gamma'\cdots\Gamma'\overrightarrow{b_kc_k}\Gamma'\overrightarrow{b'c'}$. Assume inductively that (1) holds for $\overrightarrow{b_kc_k}$, i.e. $\overrightarrow{b_ka} \in C$ and $\overrightarrow{c_ka} \in B$. Now since $\overrightarrow{b_kc_k}\Gamma'\overrightarrow{b'c'}$, both $(b_k, c_k)$ and $(b', c')$ belong to either $E_1$ or $E_2$ (or both). Further $\overrightarrow{b_k c_k} \Gamma \overrightarrow{b'c'}$. Assume without loss of generality that $(b_k, c_k), (b', c') \in E_1$. Since $a \in X \subseteq V_1$, we have $\{b_k, b', c_k, c', a\} \subseteq V_1$. Let $A_1, B_1$ and $C_1$ be the implication classes of $G_1$ such that $\overrightarrow{b_k c_k} \in A_1$, $\overrightarrow{c_k a} \in B_1$, and $\overrightarrow{b_k a} \in C_1$. Note that $A_1 \subseteq A$, $B_1 \subseteq B$ and $C_1 \subseteq C$. Since $\overrightarrow{b_k c_k} \Gamma \overrightarrow{b'c'}$, we have $\overrightarrow{b'c'} \in A_1$. Now applying Lemma 3 on triangle $ab_k c_k$ and the edge $\overrightarrow{b'c'}$, we conclude that $\overrightarrow{c'a} \in B_1$ and $\overrightarrow{b'a} \in C_1$. Therefore $\overrightarrow{c'a} \in B$ and $\overrightarrow{b'a} \in C$.

The second part of the Lemma follows directly as a consequence of the first (and noting that $A \neq C$ and $A \neq B^{-1}$). □

**Lemma 7.** *Let the vertices $a, b, c$ form a triangle in $G$ and let the edges $\overrightarrow{bc}, \overrightarrow{ca}$ and $\overrightarrow{ba}$ belong to composite classes $A, B$ and $C$ respectively with $A \neq C$, $A \neq B^{-1}$ and $B \neq C$. If $B$ and $C$ are both super classes then there exists a triangle $a', b', c'$ in $X$ with $\overrightarrow{b'c'} \in A$, $\overrightarrow{c'a'} \in B$ and $\overrightarrow{b'a'} \in C$ and hence in particular $A$ is a super class.*

*Proof.* We can assume without loss of generality that $a, b, c \in V_1$. Let the edges $\overrightarrow{ca}$ and $\overrightarrow{ba}$ belong to implication classes $I_b$ and $I_c$ (respectively) of $G_1$ (thus $I_b \subseteq B$, $I_c \subseteq C$). Hence $I_b \cap E(\hat{X}) \neq \emptyset$, $I_c \cap E(\hat{X}) \neq \emptyset$. Let $\overrightarrow{b'a''} \in I_c \cap E(\hat{X})$ and $\overrightarrow{c'a} \in I_b \cap E(\hat{X})$. Applying Lemma 3 on triangle $c, a, b$ and the edge $\overrightarrow{b'a''}$ (and noting that $A \neq C$, $A \neq B^{-1}$), we infer that $\overrightarrow{ca''} \in I_b$ and $\overrightarrow{b'c} \in A$. Now applying Lemma 3 again on triangle $b', c, a''$ and the edge $\overrightarrow{c'a'}$ (and noting that $B \neq C$ and $B \neq A^{-1}$), we infer that $\overrightarrow{b'a'} \in I_c$ and $\overrightarrow{b'c'} \in A$. This in turn implies that $A$ is a super class (since $b', c' \in X$). □

**Lemma 8.** *Let $A$ be a composite class of a graph $G = G_1 \cup G_2$. If $A \cap A^{-1} = \emptyset$, then $A \cap \hat{E}_1$ and $A \cap \hat{E}_2$ are both transitive and hence $A$ is pseudo-transitive.*

*Proof.* If $A$ is a base class then $A$ is transitive by Lemma 4. Thus we can assume that $A$ is a super class. If $A$ doesn't satisfy the conclusion of the Lemma then we can assume without loss of generality that there exist vertices $a, b, c \in V_1$ such that $\overrightarrow{ba} \in A$, $\overrightarrow{ac} \in A$ and $\overrightarrow{bc} \notin A$. If the edge $(b, c)$ is not present in $E_1$, then $\overrightarrow{ba} \Gamma' \overrightarrow{ca}$ and thus $\overrightarrow{ca} \in A \cap A^{-1}$. Therefore we can assume that $(b, c) \in E_1$. Let $C_a$ be the composite class that contains $\overrightarrow{bc}$. Now in triangle $abc$, we have $\overrightarrow{ba} \in A$ and $\overrightarrow{ca} \in A^{-1}$. Also both $A$ and $A^{-1}$ are superclasses with $A \neq C_a$. Therefore by Lemma 7, $C_a$ must be a super class, further there exists a triangle $a'b'c'$ in $X$ with $\overrightarrow{b'a'} \in A$, $\overrightarrow{a'c'} \in A$ and $\overrightarrow{b'c'} \in C_a$. But by the second condition of Lemma 6 (on triangle $b'c'a'$), $b'$ cannot be incident with an $A$ edge, a contradiction. Thus $C_a = A$ and we conclude that $A$ is pseudo-transitive. □

**Corollory 1.** *Let $A$ be a composite class of a graph $G = G_1 \cup G_2$. Then $A$ is pseudo-transitive iff $A \cap A^{-1} = \emptyset$.*

Recall that our approach involves deleting a composite class $A$ from $G$. Any composite class of $G - A$ is a union of composite classes of $G$ formed by successive "merging". Two composite classes $B$ and $C$ of $G$ are said to be *merged* by the deletion of class $A$, if deleting $A$ creates a ($\Gamma'$) relation between a $B$-edge and a $C$-edge. Note that for this to happen there must exist a triangle $a, b, c$ in $G$ with $(b, c) \in \hat{A}$ and either $\overrightarrow{ba} \in C$ and $\overrightarrow{ca} \in B$ or $\overrightarrow{ab} \in C$ and $\overrightarrow{ac} \in B$.

**Lemma 9.** *If the composite classes of $G = G_1 \cup G_2$ are all pseudo-transitive and $A$ is a base class of $G$ then the composite classes of $G - A$ are also pseudo-transitive.*

*Proof.* Let $C$ be any (composite) class of $G - A$. If $C$ is also a composite class of $G$ then it is pseudo-transitive by assumption. So assume that $C$ is formed by merging two or more composite classes of $G$.

**Claim 1.** *If base class $B_1$ merges with another composite class $M$ then $B_1 \neq M^{-1}$ and $B_1$ does not merge with any other class.*

*Proof.* Since $C$ contains a merge of $B_1$ and $M$, there exists a triangle $a, b, c$ in $G$ such that one of the following conditions hold:

1. $\overrightarrow{ba} \in M$, $\overrightarrow{ca} \in B_1$ and $\overrightarrow{bc} \in A$.
2. $\overrightarrow{ba} \in M$, $\overrightarrow{ca} \in B_1$ and $\overrightarrow{cb} \in A$.
3. $\overrightarrow{ab} \in M$, $\overrightarrow{ac} \in B_1$ and $\overrightarrow{bc} \in A$.
4. $\overrightarrow{ab} \in M$, $\overrightarrow{ac} \in B_1$ and $\overrightarrow{cb} \in A$.

All these cases are symmetric and hence we assume without loss of generality that condition (1) holds. If $B_1 = M^{-1}$ then by Lemma 8, $A = M$, a contradiction. Let $B_2 \subseteq M$ be the implication class containing the edge $\overrightarrow{ba}$. Now it is enough to show that deleting $A$ wouldn't merge $B_1$ with some other class $D \neq B_2$ of $G$. To see this, suppose deleting $A$ merges $B_1$ with $D$. Then there exists an edge $\overrightarrow{b'c'} \in A$ which together with a $B_1$ edge and a $D$ edge forms a triangle $T$ in $G$. Note that by Lemma 3, $\overrightarrow{b'a} \in B_2$ and $\overrightarrow{c'a} \in B_1$. Therefore the $B_1$ edge of $T$ must be $\overrightarrow{c'a'}$ for some vertex $a'$. (Applying the second part of Lemma 3 on $b'c'a$ we infer that $b'$ cannot be adjacent to a $B_1$ edge. Also by the same Lemma (applied on $b'c'a$ and $\overrightarrow{a'c'}$), if $\overrightarrow{a'c'} \in B_1$ then $\overrightarrow{b'c'} \in B_2$, a contradiction). Hence $\overrightarrow{b'a'}$ is the $D$-edge of $T$. Now applying Lemma 3 again on $b'c'a$ and edge $\overrightarrow{c'a'}$, we infer that $D = B_2$. □

**Claim 2.** *Two super classes don't merge in $C$.*

*Proof.* Assume (for the sake of contradiction) that two super classes $S_y$ and $S_z$ merge in $C$. Thus we can assume without loss of generality that there exists a triangle $a, y, z$ in $G$ with $\overrightarrow{yz} \in A$, $\overrightarrow{za} \in S_y$ and $\overrightarrow{ya} \in S_z$. (The other cases are symmetric as observed in Claim 1). Since $A$ is disjoint from $\hat{S}_y$ and $\hat{S}_z$ and $S_y \neq S_z$, the conditions of Lemma 7 are satisfied. Thus applying Lemma 7 on triangle $ayz$, we conclude that $A$ is a super class, a contradiction. □

**Claim 3.** *If $C$ contains a super class say, $S$ (of $G$), then $C$ is of the form $C = S \cup B_1, \cdots, B_k$, where $B_1, \cdots, B_k$ are the base classes of $G$ and $B_i \neq B_j^{-1}$ for $1 \leq i, j \leq k$. Otherwise $C$ is the union of two base classes $B_1$ and $B_2$ with $B_1 \neq B_2^{-1}$.*

*Proof.* This follows directly as a consequence of Claims 1 and 2. □

Claim 3 implies that $C \cap C^{-1} = \emptyset$ and hence $C$ is pseudo-transitive (by Lemma 8). □

Our method will be to remove base classes first. The following Lemma examines what happens when all the composite classes are super classes and one of them gets deleted.

**Lemma 10.** *Let each of the composite classes of $G = G_1 \cup G_2$ be super and pseudo-transitive. If $A$ is any super class of $G$ then each of the composite classes of $G - A$ is pseudo-transitive.*

*Proof.* Let $L$ be a composite class of $G - A$. If $L$ is also a composite class of $G$ then $L$ is pseudo-transitive by assumption. So assume that $L$ is not a composite class of $G$. We claim that $L$ consists of precisely a merge of two super classes. To see this let $L$ contains the merge of super classes $B$ and $C$. We can assume (without loss of generality) that there exists a triangle $abc$ in $G$ with $\overrightarrow{bc} \in A$, $\overrightarrow{ca} \in B$ and $\overrightarrow{ba} \in C$. Further by Lemma 7, we can assume that $\{a, b, c\} \in X$.

We now show that deleting $A$ wouldn't merge $B$ with some other super class $D \neq C$. The proof is parallel to that of Claim 1, though we can't appeal to Lemma 3 as we do there, and must use Lemma 6 instead. If $L$ contains a merge of $B$ with some other super class $D$, then there exists a triangle $T = a'b'c'$ in $G$ with $\overrightarrow{b'c'} \in A$ and the other two edges in $B$ and $D$. Further by Lemma 7, we can assume that $\{a', b', c'\} \subseteq X$. Note that by Lemma 6, $\overrightarrow{b'a} \in C$ and $\overrightarrow{c'a} \in B$. Therefore the $B$ edge of $T$ must be $\overrightarrow{c'a'}$. (Applying the second part of Lemma 6 on $b'c'a$ we infer that $b'$ cannot be adjacent to a $B$ edge. Also by the same Lemma applied on $b'c'a$ and $\overrightarrow{a'c'}$, if $\overrightarrow{a'c'} \in B$ then $\overrightarrow{b'c'} \in C$, a contradiction). Hence $\overrightarrow{b'a'}$ is a $D$-edge of $T$. Now applying Lemma 6 again on $b'c'a$ and the edge $\overrightarrow{c'a'}$, we infer that $D = C$.

Therefore $B$ doesn't merge with any class other than $C$ and similarly $C$ doesn't merge with any class other than $B$. Hence $L$ consists of precisely a merge of two super classes $B$ and $C$ and therefore $L$ is pseudo-transitive (since $B \neq C^{-1}$). $\square$

A partition of the edge set $\hat{E}_1 \cup \hat{E}_2 = \hat{B}_1 + \hat{B}_2 + \cdots + \hat{B}_i + \hat{S}_{i+1} + \hat{S}_{i+2} + \cdots + \hat{S}_j$ is said to be a $S$-decomposition of $G = G_1 \cup G_2$, if $B_k$ is a base class of $G - \cup_{1 \leq l < k} \hat{B}_l$ and $S_k$ is a super class of $G - \cup_{1 \leq l \leq i} \hat{B}_l - \cup_{i+1 \leq l < k} \hat{S}_l$

We are now ready to prove the main theorem.

**Theorem 5.** *Let $G_1 = (V_1, E_1)$ and $G_2 = (V_2, E_2)$ be two comparability graphs sharing some vertices $X$ and the edges induced by $X$. Let $\hat{E}_1 \cup \hat{E}_2 = \hat{B}_1 + \hat{B}_2 + \cdots + \hat{B}_i + \hat{S}_{i+1} + \hat{S}_{i+2} + \cdots + \hat{S}_j$ be a $S$-decomposition of $G_1 \cup G_2$. The following statements are equivalent.*

1. *$G_1$ and $G_2$ are simultaneous comparability graphs*

2. *Every composite class of $G = G_1 \cup G_2$ is pseudo-transitive, i.e. $C \cap C^{-1} = \emptyset$ for all composite classes $C$ of $G = G_1 \cup G_2$.*

3. *Every partition of the $S$-decomposition is pseudo-transitive, i.e. $B_k \cap B_k^{-1} = \emptyset$ for $k = 1, \cdots, i$ and $S_k \cap S_k^{-1} = \emptyset$ for $k = i+1, \cdots, j$.*

*Proof.* (1) $\Rightarrow$ (2) follows from Lemmas 5 and 8.

(2) $\Rightarrow$ (3) is a direct consequence of Lemmas 9 and 10.

(3) $\Rightarrow$ (1)

Let $T = B_1 + B_2 + \cdots + B_i + \cdots S_{i+1} + S_{i+2} + \cdots S_j$. We now claim that $T$ is pseudo-transitive. For $k = 1, \cdots, j$, define $C_k$ as $C_k = B_k$ if $k \leq i$ and $C_k = S_k$ otherwise. Thus $T = C_1 + \cdots C_j$.

For $k = 1 \cdots j$, let $T_k = C_k + \cdots C_j$. (Thus $T_1 = T$) and $H_k = \hat{C}_k + \cdots \hat{C}_j$. Thus $C_k$ is a composite class of $H_k$. Now it is enough to show that $T_k$ is pseudo-transitive for any $k$. Assume inductively that $T_{k+1} = T_k - C_k$ is pseudo-transitive. Note that $\hat{T}_{k+1} \cap \hat{C}_k = \emptyset$. Now we claim that $T_k = T_{k+1} \cup C_k$ is also pseudo-transitive.

Suppose not. Then there exist vertices $a, b, c$ all in $G_1$ or all in $G_2$ such that $\overrightarrow{ab} \in T_k$, $\overrightarrow{bc} \in T_k$ and $\overrightarrow{ac} \notin T_k$. Since $T_{k+1}$ and $C_k$ are pseudo-transitive we only have to consider the case when $\overrightarrow{ab} \in T_{k+1}$ and $\overrightarrow{bc} \in C_k$ (the other case $\overrightarrow{ab} \in C_k$ and $\overrightarrow{bc} \in T_{k+1}$ is symmetric).

Now if the edge $(a, c)$ is not present in $H_k$ then $\vec{bc}\Gamma'\vec{ba}$ and thus $\vec{ba} \in C_k$, contradicting that $\hat{T}_{k+1} \cap \hat{C}_k = \emptyset$. So either $\vec{ca} \in T_{k+1}$ or $\vec{ca} \in C_k$. This implies (by the pseudo-transitivity of $T_{k+1}$ and $C_k$) that $\vec{cb} \in T_{k+1}$ or $\vec{ba} \in C_k$. In both cases we get a contradiction to $\hat{T}_{k+1} \cap \hat{C}_k = \emptyset$.

Thus all four cases lead to a contradiction and we conclude that $T_k$ is pseudo-transitive. □

Theorem 5 gives rise to the following $O(nm)$ algorithm for recognizing simultaneous comparability graphs: Given graphs $G_1$ and $G_2$ check whether all composite classes of $G = G_1 \cup G_2$ are pseudo-transitive. If so return YES otherwise return NO.

If $G_1$ and $G_2$ are simultaneous comparability graphs then the following algorithm computes an S-decomposition of $G_1 \cup G_2$. As shown in the proof of Theorem 5 this immediately gives a pseudo-transitive orientation.

**Algorithm 2**
1. Initialize $T = \emptyset$.
2. Compute all base-classes $B_1, B_1^{-1}, \cdots, B_b, B_b^{-1}$.
3. **For** $i = 1$ to $b$, place all the edges of $B_i$ (resp $B_i^{-1}$) in a separate set labelled $i$ (resp $-i$).
4. Place all the remaining edges in one set and assign a label 0.
5. Let $S_{\vec{uv}}$ denote the set containing $\vec{uv}$.
6. **While** there exists a set $S$ with non-zero label **Do**:
7.   Add all (directed) edges of $S$ to $T$.
8.   Assign label 0 to $S$ and $S^{-1}$.
9.   **For** each edge $\vec{bc}$ in $S$ and each vertex $a$ in $G$ such that $abc$ forms a triangle **Do**:
10.     **If** labels of $S_{\vec{ab}}$ and $S_{\vec{ac}}$ are equal: **Continue**.
11.     Let $l$ be a label defined as: $l = 0$ if labels of $S_{\vec{ab}}$ or $S_{\vec{ac}}$ is 0, otherwise $l =$ label of $S_{\vec{ab}}$.
12.     Merge $S_{\vec{ab}}$ and $S_{\vec{ac}}$ and assign label $l$ to the union.
13.     Merge $S_{\vec{ba}}$ and $S_{\vec{ca}}$ and assign label $-l$ to the union.
14.   **End**
15. **End**
/* Now each composite class of $G' = G - \hat{T}$ is a super class. */
16. **While** $G'$ is non-empty **Do**:
17.   Let $C$ be a super class of $G'$.
18.   $T = T \cup C$ and $G' = G' - \hat{C}$.
19. **End**
20. **Return** $T$.

Given two simultaneous graphs $G_1$ and $G_2$, Algorithm 2 computes the pseudo-transitive orientation of $G_1 \cup G_2$ as follows. We first compute all the base classes and distinguish them from super classes using labels (lines 2 3 and 4). We then iteratively find a base class, add it to the solution and delete it (lines 5-8). By Lemma 9 deletion of a base class may leave other composite classes unchanged, or merge two base classes or merge a super class with a set of base classes. We handle these cases in lines 10 to 13 and update the labels. After the while loop terminates, we are only left with super classes. These are handled in lines 16 to 19.

Algorithm 2 can be implemented to run in $O(nm)$ time using the standard disjoint-set data structures. We do at most $O(m)$ unions and $O(nm)$ finds. Hence for the complexity analysis, we assume that the (amortized) run time of each set operation is $O(1)$. Now consider the run time of each of the steps: Computing all the composite classes (base and super) takes $O(nm)$ time. Thus line 2 takes $O(nm)$ time.

Lines 3 and 4 take $O(m)$ time. In line 6, finding a set with non-zero label takes at most $O(m)$ time. In each iteration of the while loop, if the chosen set has $m_1$ elements, then the For loop (lines 9-14) takes $O(m_1 n)$ time. Hence the total run time of the while loop is $O((m_1 + m_2 + \cdots m_i)n)$ where $m_i$ is the size of the set chosen in the $i$th iteration of the algorithm. This in turn is at most $O(nm)$. Lines 16-20 also run in $O(nm)$ time. Hence the run time of Algorithm 2 is $O(nm)$.

**Remark:** Note that if $T$ is a pseudo-transitive of $G_1 \cup G_2$, then $T$ can be augmented to a transitive orientation by computing $T' = T^2$ (as shown in the proof of Theorem 3. The complexity of this step is same as the complexity of matrix multiplication: $O(n^{2.376})$. Hence computing an augmented comparability graph takes $O(nm + n^{2.376})$ steps.

## 4 Simultaneous Permutation Graphs

A graph $G = (V, E)$ on vertices $V = \{1, \cdots, n\}$ is said to be a *permutation graph* if there exists a permutation $\pi$ of the numbers $1, 2, \cdots, n$ such that for all $1 \leq i < j \leq n$, $(i, j) \in E$ if and only if $\pi(i) > \pi(j)$. Equivalently, $G = (V, E)$ is a permutation graph if and only if there are two orderings $L$ and $P$ of $V$ such that $(u, v) \in E$ iff $u$ and $v$ appear in the opposite order in $L$ and in $P$. We call $\langle L, P \rangle$ an *order-pair* for $G$. The intersection representation for permutation graphs follows immediately: $G = (V, E)$ is a permutation graph iff there are two parallel lines $l$ and $p$ and a set of line segments each connecting a distinct point on $l$ with a distinct point on $p$ such that $G$ is the intersection graph of the line segments. Observe that $L$ and $P$ correspond to the ordering of the endpoints of the line segments on $l$ and $p$ respectively. Since permutation graphs are a class of intersection graphs the equivalence theorem 1 is applicable for this class.

Let $G_1 = (V_1, E_1)$ and $G_2 = (V_2, E_2)$ be two permutation graphs sharing some common vertices $X$ and the edges induced by $X$.

By the definition of simultaneous intersection graphs, $G_1$ and $G_2$ are simultaneous permutation graphs iff there exist two parallel lines $l$ and $p$ and a set of line segments joining distinct points on $l$ and $p$ such that the vertices $V_1 \cup V_2$ are in one-to-one correspondence with the line segments and such that for $i = 1, 2$ and for $u, v \in V_i$, we have $(u, v) \in E_i$ iff the corresponding line segments intersect. We begin with a "relaxed" characterization of simultaneous permutation graphs in terms of order-pairs.

**Lemma 11.** $G_1$ and $G_2$ are simultaneous permutation graphs iff there exist order-pairs $\langle L_1, P_1 \rangle$ and $\langle L_2, P_2 \rangle$ for $G_1$ and $G_2$ (respectively) such that every pair of vertices $u, v \in X$ appear in the same order in $L_1$ as in $L_2$ AND appear in the same order in $P_1$ as in $P_2$.

*Proof.* The forward direction is clear. For the reverse direction, we create a total order $L$ on $V_1 \cup V_2$ consistent with both $L_1$ and $L_2$. This is possible because $L_1$ and $L_2$ are consistent on $X$. We do the same for $P$. The orderings $L$ and $P$ provide the endpoints of line segments for the simultaneous intersection representations of $G_1$ and $G_2$. □

We make use of the following characterization of permutation graphs.

**Theorem 6.** *(Pnueli, Lempel and Even [7]).* $G$ is a permutation graph if and only if $G$ and its complement $\bar{G}$ are both comparability graphs.

We prove an analogous result for simultaneous permutation graphs.

**Theorem 7.** *Let $G_1 = (V_1, E_1)$ and $G_2 = (V_2, E_2)$ be two undirected graphs sharing some vertices $X$, and the edges induced by $X$. $G_1$ and $G_2$ are simultaneous permutation graphs if and only if $G_1$ and $G_2$ are simultaneous comparability graphs and simultaneous co-comparability graphs.*

*Proof.* Let $G_1$ and $G_2$ be simultaneous permutation graphs. By Theorem 1 there exists an augmenting set of edges $A \subseteq (V_1 - X) \times (V_2 - X)$, such that $G = (G_1, G_2, A)$ is a permutation graph. Now by Theorem 6, $G$ and $\bar{G} = (\bar{G}_1, \bar{G}_2, (V_1 - X) \times (V_2 - X) - A)$ are comparability graphs. Hence by Theorem 3, $G_1$ and $G_2$ are simultaneous comparability graphs and also $\bar{G}_1$ and $\bar{G}_2$ are simultaneous comparability graphs.

For the other direction, let $G_1$ and $G_2$ be simultaneous comparability graphs and $\bar{G}_1$ and $\bar{G}_2$ also be simultaneous comparability graphs. Let $F_1$ and $F_2$ be the transitive orientations of $G_1$ and $G_2$ that are consistent on the edges of $G_1 \cap G_2$. Also let $R_1$ and $R_2$ be the transitive orientations of $\bar{G}_1$ and $\bar{G}_2$ that are consistent on the edges of $\bar{G}_1 \cap \bar{G}_2$. As shown in [7], $F_1 + R_1$ and $F_1^{-1} + R_1$ are acyclic transitive orientations of $G_1$. Following the original idea of Pneuli et al [7], we define an order-pair $\langle L_1, R_1 \rangle$ on $V_1$ as follows: let $L_1$ be a total order of $V_1$ consistent with the partial order $F_1 + R_1$; and let $P_1$ be a total order of $V_1$ consistent with the partial order $F_1^{-1} + R_1$. Similarly define $\langle L_2, R_2 \rangle$ on $V_2$ using $F_2 + R_2$ and $F_2^{-1} + R_2$.

We now show that any two vertices $u, v \in X$ satisfy the conditions of Lemma 11.

*Case 1.* $(u, v) \in E_1$ (and hence $E_2$): Without loss of generality assume that the edge is directed from $u$ to $v$ in $F_1$ and $F_2$. Note that $L_1(u) < L_1(v)$ and $L_2(u) < L_2(v)$ (since $F_1 + R_1$ and $F_2 + R_2$ are transitive orientations). Similarly $P_1(u) > P_1(v)$ and $P_2(u) > P_2(v)$.

*Case 2.* $(u, v) \in \bar{E}_1$ (and hence $\bar{E}_2$): Without loss of generality assume that the edge is directed from $u$ to $v$ in $R_1$ and $R_2$. We have $L_1(u) < L_1(v)$, $L_2(u) < L_2(v)$, $P_1(u) < P_1(v)$ and $P_2(u) < P_2(v)$.

From the above two cases, the conditions of Lemma 11 hold true for $G_1$ and $G_2$ and hence we conclude that $G_1$ and $G_2$ are simultaneous permutation graphs. □

Since simultaneous comparability graphs can be recognized in $O(nm)$ time, Theorem 7 implies that simultaneous permutation graphs can be recognized in $O(n^3)$ time. We also note that a similar approach was used in [5] to recognize probe permutation graphs.

## 5 Discussion

A main contribution of this paper is the introduction of the simultaneous membership problem, which is closely related to the probe graph recognition and graph sandwich problems. We gave poly-time algorithms for solving the problem for chordal, comparability and permutation graphs. The running time of our algorithm for comparability graphs is $O(nm)$, which matches the run time of the best-known algorithm for (partitioned) probe comparability graphs [5]. For chordal and permutation graphs both of our algorithms run in $O(n^3)$. (The best-known algorithms for probe chordal and probe permutation graphs are $O(nm)$ [10] and $O(n^2)$ [4] respectively).

We believe that the simultaneous membership problem for interval graphs is also solvable in polynomial time, but it seems substantially more difficult than for chordal and comparability graphs. We have a solution in progress [13].

It would be interesting to unify our recognition algorithms for simultaneous chordal and comparability graphs with the recent recognition algorithms for probe chordal and probe comparability graphs respectively. The obvious generalization is to have a set of augmenting edges of the form $X \times Y$, where $X$ and $Y$ are subsets of vertices. Probe graphs are the case $X = Y$, giving a clique of augmenting edges, and simultaneous graphs are the case where $X$ and $Y$ are disjoint, giving a complete bipartite graph of augmenting edges.